\newcommand\sovast{\ref@jnl{Soviet~Ast.}} 

\documentclass[12pt,usenatbib]{mn2e}
\usepackage{graphicx}
\usepackage{epsf}
\usepackage{bm}
\usepackage{color}
\usepackage{rotating}
\usepackage[T1]{fontenc}
\usepackage{ae,aecompl}

\def\lsim{\mathrel{\lower0.6ex\hbox{$\buildrel {\textstyle <}
 \over {\scriptstyle \sim}$}}}
\def\gsim{\mathrel{\lower0.6ex\hbox{$\buildrel {\textstyle >}
 \over {\scriptstyle \sim}$}}}

\def\hmpc{ h^{-1}\rm Mpc}

\def\eone{${\bf e}_{1}$}
\def\etwo{${\bf e}_{2}$}
\def\ethree{${\bf e}_{3}$}

\def\rnl{$r_{\rm NL}$}

\def\rs{$r_{\rm s}$}
\def\Rs{$R_{\rm s}$}

\begin{document}

\title[The velocity shear and vorticity ]{The velocity shear and vorticity across redshifts and non-linear scales
}
\author[Libeskind et al.] 
{Noam I Libeskind$^{1}$, Yehuda Hoffman$^2$ \& Stefan Gottl\"ober$^1$\\
 $^1$Leibniz-Institute f\"ur Astrophysik Potsdam (AIP), An der Sternwarte 16, D-14482 Potsdam, Germany\\
 $^2$Racah Institute of Physics, Hebrew University, Jerusalem 91904, Israel\\
  }
\date{Accepted --- . Received ---; in original form ---}
\pagerange{\pageref{firstpage}--\pageref{lastpage}} \pubyear{2002}
\maketitle

 \begin{abstract}
The evolution of the large scale distribution of matter in the universe is often characterized by the density field. 
Here we take a complimentary approach and characterize it using the cosmic velocity field, specifically the deformation of the velocity field. The deformation tensor is decomposed 
into its symmetric component (known as the ``shear tensor'') and its anti-symmetric part (the ``vorticity''). Using a high resolution cosmological simulation we examine the relative orientations of the shear and the vorticity as a function of spatial scale and redshift. The shear is found to be remarkable stable to the choice of scale, while the vorticity is found to quickly decay with increasing spatial scale or redshift. 
The vorticity emerges out of the linear regime randomly oriented with respect to the shear eigenvectors. 
Non-linear evolution drives the vorticity to lie within the plane defined by the eigenvector of the fastest collapse. Within that plane the vorticity first gets aligned with the middle eigenvector and then it moves to be preferentially aligned with the third eigenvector, of slowest collapse.  Finally a scale of ``non-linearity'' to be used when calculating properties of the non-linear deformation tensor at different redshifts is suggested.

\end{abstract}

\section{Introduction}
\label{section:intro} 

In the standard model of cosmology, structure grows primarily via gravitational instability in an expanding universe. Accordingly, the expansion is homogeneous and isotropic, up to some small linear primordial perturbations that were created in an early inflationary phase \citep[e.g.][]{1983PhRvD..28..679B}. Gravitational instability, induced by the primordial perturbations, drives the dark matter (DM) component to cluster and collapse into virial halos \citep{1978MNRAS.183..341W}, forming a network of voids, sheets, filaments and dense knots, the so-called the cosmic web \citep{1996Natur.380..603B}. Gravitational instability is often treated analytically in the linear and quasi-linear phases of structure formation \citep{1970A&A.....5...84Z,1980lssu.book.....P}. The fully non-linear regime is virtually intractable analytically and has led the community to rely heavily on numerical simulations of large scale structure, halo clustering and galaxy formation \citep[][to name just a few]{1979ApJ...228..664A,1983ApJ...271..417F,2006Natur.440.1137S}.

The emergence of structure implies a change to the density field in time and, by the equation of continuity, this induces a time-dependent velocity field. It follows that both the density and velocity fields are involved in the gravitational dynamics \citep[e.g.][]{1980lssu.book.....P}. It is customary to describes the full dynamical evolution by means and terms involving the density field. This 
des the correlation function of the mass and galaxy distribution, moments of the density field, distribution and mass functions of halos as well as various density-based measures of the cosmic web \citep[for example][]{Sousbieetal2008,2010PhRvD..81j3006S}.

However, the growth of structure can  equally well be described by means of the velocity field. The motivation for describing cosmology by means of the density is clear - the distribution of galaxies, and by implication of DM, is observationally much more accessible than their velocities (by, for example, large aperture sky surveys). Yet, it is the aim of the present paper to follow the dynamics of some measures of the velocity field across different spatial and temporal scales.

The cosmic web constitutes one of the most conspicuous characteristics of the large scale structure (LSS) of the universe. This is the unequivocal impression one gets from visual inspection of the distribution of galaxies in the actual universe and in numerical simulations. The translation of this visual impression into a rigorous mathematical description is not trivial and many algorithms for classifying the cosmic web have been proposed  \citep[e.g.][see also \citealt{2013MNRAS.429.1286C} for a comparison of some of these methods]{2005A&A...434..423S,2006ApJ...645..783S,2007ApJ...655L...5A,Hahnetal2007a,Sousbieetal2008,Forero-Romero2009,2010ApJ...723..364A,2012MNRAS.425.2049H}.
A  robust and computationally efficient algorithm for the construction of the cosmic web in numerical simulations has been recently presented \citep[hereafter the V-web;][]{2012MNRAS.425.2049H,2012MNRAS.421L.137L,2013MNRAS.428.2489L}. This is based on the analysis of the local velocity shear tensor, namely the symmetric deformation tensor constructed from the spatial derivatives of the velocity field. 

One of the prime motivations for studying the cosmic web is the desire to understand the formation and properties of the DM halos (and therefore of galaxies) that inhabit it. For example, the cosmic web may play an important role in the accretion of satellites \citep{Knebeetal2004,Libeskindetal2005,Libeskindetal2011a}. The alignment of DM halo spin with respect to the cosmic web strongly suggests an important relationship between DM halos and the cosmic web. This has led to a plethora of papers dealing with the various aspects of the alignment in simulations \citep[][among others]{Hahnetal2007a,Aragon-Calvoetal2007,2007MNRAS.375..184B,Sousbieetal2008,2013ApJ...766L..15L,2012MNRAS.421L.137L}  as well as observations \citep{LeeErdogdu2007,2001ApJ...555..106L,2013ApJ...775L..42T}

That a halo spin alignment can be very easily measured and calculated in simulations, is often taken for granted, despite the non-negligible obstacles that all the above studies face; namely the fact that both DM halos and the cosmic web are not robustly and uniquely defined objects. Even DM halos, that are more simply conceived than the cosmic web, are not uniquely defined \citep[see][for an excellent review of differences in halo finders used by the community]{2013MNRAS.435.1618K}.
The diversity in the available definitions of the cosmic web in someways reflects its nebulous, vague nature. This has encouraged us to adopt a new approach to the problem, which bypasses some of these difficulties. 
\begin{figure}
 \includegraphics[width=20pc]{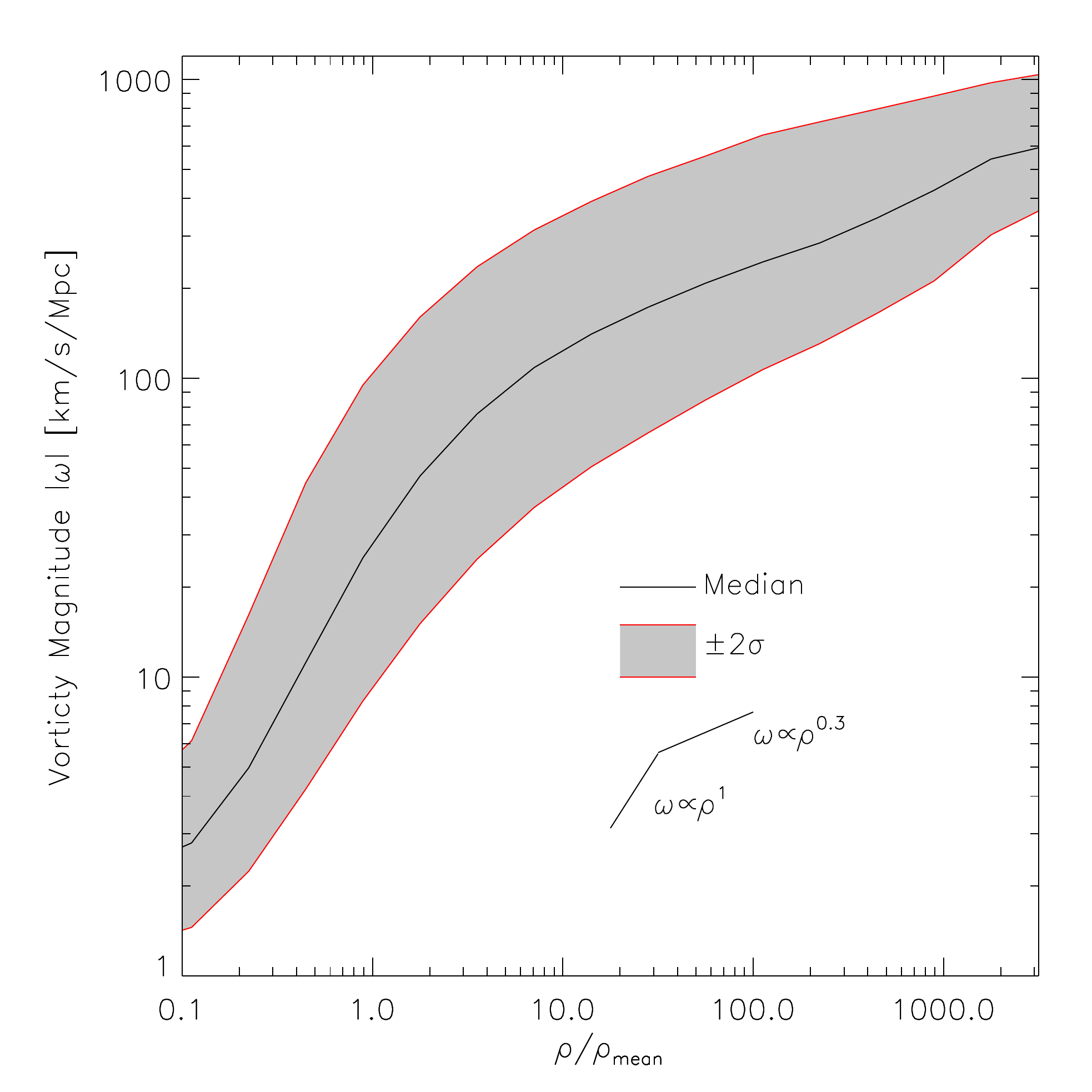}
 \caption{The magnitude of the vorticity, $|\omega|$ as a function of density measured in units of the mean (at $z=0$). Shown here is the median (black line) and the $\pm2\sigma$ spread in vorticity magnitude. The vorticity is a strong function of density, and decays quickly in low density environments. }
\label{fig:magvortdens}
 \end{figure}

\begin{figure*}
 \includegraphics[width=40pc]{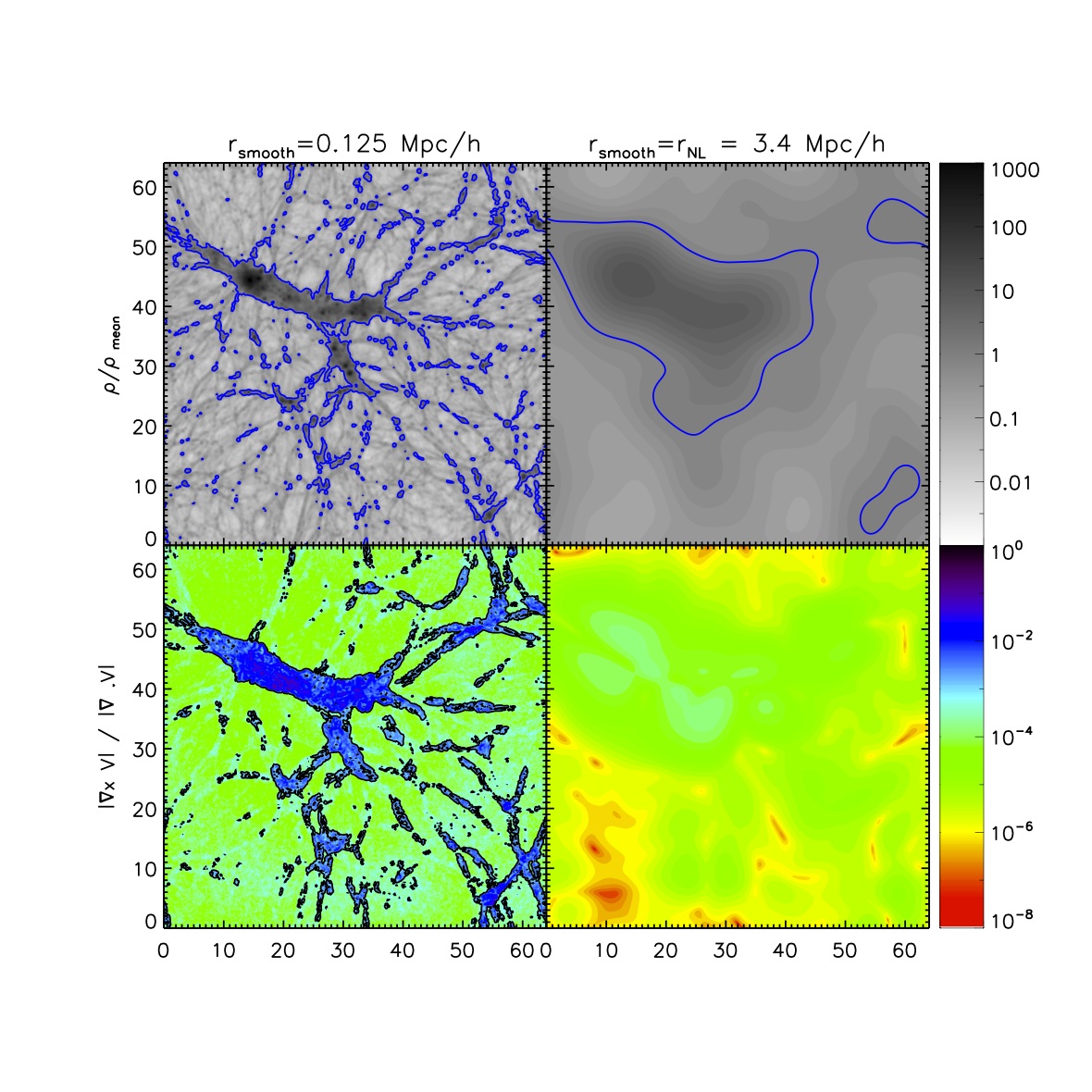}
 \caption{The $z=0$ density (top) and vorticity (bottom) fields in a 0.125$h^{-1}$Mpc slice of the simulation box, smoothed with $r_{\rm s}=0.125h^{-1}$Mpc and 3.4$h^{-1}$Mpc on the left and right, respectively. \textit{Top Panels:} The density field is shown in units of the mean and logarithmically contoured. The blue contour denotes the mean density. \textit{Bottom Panels:} The vorticity is measured relative to the divergence of the velocity and logarithmically contoured. The black line denotes regions where the vorticity equals 1\% of the divergence of the velocity. No black line is present  when smoothed with 3.4$h^{-1}$Mpc since on these scales the vorticity never grows to be greater than 1\% of the velocity.}
 \label{figure:maps}
 \end{figure*}

\cite{2013ApJ...766L..15L} analyzed the role of the shear tensor and the vorticity in shaping the cosmic web and the alignment of the spin of DM halos. Indeed the importance of the vorticity was examined in a seminal work by \cite{2009PhRvD..80d3504P} who used a large cosmological simulation to compute  the vorticity (and velocity divergence) power spectrum. Among other important contributions, \cite{2009PhRvD..80d3504P} showed that although the vorticity is (as expected) negligible on linear scales, it can still be non-zero there due to noise in the way it is numerically computed. Thus care has to be taken when computing vorticity from cosmological simulations.

The overall picture that emerges is that the symmetric and antisymmetric parts of the deformation tensor have a major role in shaping the cosmic web and determining the principal directions of haloes within the framework of the web. The two main results of \cite{2013ApJ...766L..15L} were: a. The spin of DM halos is strongly aligned with the vorticity of the velocity field; b. The eigenvectors of the velocity shear tensor determine the direction of DM halo spin. The nature of the spin-vorticity-shear relation is independent of the web classification of the halos, but its strength shows a mild dependence on it. Namely, the 
spins are determined by the unambiguous and very clearly defined principal directions of the shear tensor. That the spin is aligned with the somewhat vaguely defined cosmic web, is a corollary of the fact that the web is shaped by the shear tensor. 

\cite{2013arXiv1307.1232T} recently provided strong support to the conjecture that the web is determined by the velocity shear tensor. That paper presented a comparative analysis of two different methods of extracting filaments out of the DM distribution: the V-web and the so-called Bisous model \citep[a marked point process with interactions, see][]{Stoica:05,2005A&A...434..423S}. \cite{2013arXiv1307.1232T} find that that in spite of the considerable lack of overlap between the two classes of filaments, the Bisous filaments are strongly aligned with the eigenvector of the smallest eigenvalue of the shear tensor, as predicted by the V-web. This is not a trivial results as the construction of the Bisous has no explicit dependence on the shear tensor or the velocity field. This conjecture has therefore motivated us to analyze the orientation of the vorticity vector and the principal directions of the shear tensor across different spatial scales and cosmological times, rather than studying the orientation of the spin of halos and the cosmic web \citep[e.g.][]{2013MNRAS.428.1827T}. The former is more fundamental than the latter.

The velocity shear tensor and  the vorticity at $z=0$, were 
studied in \cite{2013ApJ...766L..15L} only and at fixed scales: time evolution and scale dependence were not examined. We turn now to  to look at the behavior of the shear and the vorticity across different temporal and spatial scales. We wish to characterize the time-dependent and multi-scale nature of the phenomena detailed in \cite{2013ApJ...766L..15L}, in particular, the preferential direction of the vorticity with respect to the local principal directions established by the eigenvectors of the shear tensor. This is done by analyzing a high resolution cosmological DM-only simulation.

The structure of the paper is as follows. Sec.~\ref{section:background} outlines the mathematical framework, which consists of interpolating the velocities of DM particles on Clouds-in-Cells (CIC) grid and evaluating the spatial derivatives of the velocity field on the CIC grid. The results of the analysis of the spatial and temporal behavior of the shear tensor and the vorticity are reported in Sec.~\ref{section:results}. 
A general summary and a discussion on the theoretical and observational implication of the present of the paper are presented in Sec.~\ref{section:summary}.

\section{Background and Method}
\label{section:background} 
In this section, a brief review of the decomposition of the peculiar velocity field is given \citep[see][for a comprehensive study of the deformation tensor]{1972fct..book.....T}. 
Consider the peculiar velocity field,  
${\bf v}({\bf r}, z)$, 
a differential distance away from some reference point ${\bf r}_{0}$. The velocity field can be Taylor expanded about ${\bf r}_{0}$ to 1st order such that 
\begin{equation}
v_{i}({\bf r})=v_{i}({\bf r}_{0})+\frac{\partial v_{i}}{\partial r_{j}}\bigg|_{{\bf r}_{0}}{\rm d}{\bf r}
\end{equation}
where $\rm d{\bf r}={\bf r-r}_{0}$ is the displacement and $i,\  j,\  k=x,\ y, \  z$ are the components. The deformation tensor can be written
\begin{equation}
\frac{\partial {v}_{i}}{\partial r_{j}} = \frac{1}{2}\left(\frac{\partial v_{i}}{\partial r_{j}}+\frac{\partial v_{j}}{\partial r_{i}}\right)+\frac{1}{2}\left(\frac{\partial v_{i}}{\partial r_{j}}-\frac{\partial v_{j}}{\partial r_{i}}\right)
\end{equation}
or expanded as the sum of a symmetric and anti-symmetric components, namely:
\begin{eqnarray}
\frac{\partial {v}_{i}}{\partial r_{j}} &=& \big( -\Sigma_{ij} + \Gamma_{ij} \big) \times H(z)
\end{eqnarray}
where
\begin{eqnarray}
\Sigma_{ij}&=&-\frac{1}{2H(z)}\bigg(\frac{\partial v_{i}}{\partial r_{j}}+\frac{\partial v_{j}}{\partial r_{i}}\bigg)\label{eq:shear}\\
\Gamma_{ij}&=&\frac{1}{2H(z)}\bigg(\frac{\partial v_{i}}{\partial r_{j}}-\frac{\partial v_{j}}{\partial r_{i}}\bigg)\label{eq:vort}. 
\end{eqnarray}
Note that the deformation tensor is scaled by the Hubble constant at a given redshift $H(z)$, where 
$H^{2}(z)=H_{0}\left(\Omega_{\rm m}(1+z)^{3}+\Omega_{\Lambda}\right)$
 such that it is dimensionless. Also, the shear tensor $\Sigma_{ij}$ has been defined with a minus sign so as to make its largest eigenvalue correspond to the fastest collapsing axis. Note that the the trace-free component of $\Sigma_{ij}$ causes a shear\footnote{The word ``strain'' or ``shear strain'' is also used fluid mechanics to denote the off diagonal elements of $\Sigma_{ij}$.} while change in volume of the deformation tensor is captured by the the trace of $\Sigma_{ij}$ which corresponds to the compression (or divergence) of the velocity field:
\begin{eqnarray}
{\rm Tr}({\Sigma}) =  -{\nabla \cdot {\bf v}\over H(z)}
\end{eqnarray}
This is proportional to the overdensity in the linear regime (i.e. $-\nabla \cdot {\bf v}\propto \delta$).  
The skew anti-symmetric part, $\Gamma_{ij}$ is often called the vorticity tensor since it is composed of elements of the vorticity vector:
\begin{equation}
\Gamma_{ij}=-\frac{1}{2 H(z) }\epsilon_{ijk}\omega_{k}
\end{equation}
where $\epsilon_{ijk}$ is the fully anti-symmetric Levi-Cevita symbol and 
${\boldsymbol \omega}=\nabla\times{\bf v}$ 
is the vorticity or curl of the velocity field. Note that the components of the vorticity vector may also be computed directly from the vorticity tensor 
\begin{equation}
\omega_{i} =\epsilon_{ijk}\Gamma_{jk} H(z)
\end{equation}
The shear tensor $\Sigma_{ij}$ is symmetric and therefore can be diagonalized to obtain its three eigenvalues (termed and ordered by convention as $\lambda_{1}>\lambda_{2}>\lambda_{3}$) and corresponding eigenvectors (\eone, \etwo, and \ethree). The eigenvectors are non-directional and thus define an orthonormal basis; often they are referred to as an eigenframe or an eigensystem.

\subsection{Simulation and CIC}

In order to characterize the evolution of the fully non-linear velocity field according to the decomposition described above, a 
DM only $N$-body simulation in the $\Lambda$CDM cosmogony of a 64$h^{-1}$Mpc box populated with $2048^{3}$ particles run by the GADGET2 code  is employed \citep{2005MNRAS.364.1105S}
\citep[see][]{2010MNRAS.401.1889L,2010MNRAS.405.1119K}\footnote{This is a constrained simulation of the local universe, however its constrained nature is not used here. See {\tt http://www.clues-project.org/}}. The simulations achieve a resolution of $\sim2\times10^{6}h^{-1}$M$_{\odot}$ per mass element and a force resolution of $1.5$kpc$/h$. The simulation adopts standard WMAP5 cosmological parameters \citep[e.g.][]{Komatsu09}.

At each snapshot the velocity field is gridded according to a ``cloud-in-cell''  (CIC) scheme. An upper limit on the size of the CIC employed is set by ensuring that every mesh cell contains at least one particle at $z=0$ (the motivation being that a mesh that meets this requirement at $z=0$ will meet it at all redshifts; indeed this is the case here). This rule implies a $512^{3}$ CIC, resulting in cells of side-length $0.125h^{-1}$Mpc. 

In order to suppress the artificial cartesian preferred directions 
introduced by the CIC, a gaussian smoothing is applied to the CIC density and velocity fields. The gaussian smoothing also sets the scale of the shear and vorticity calculation. When a small smoothing kernel is used, the highly non-linear features of the velocity field are apparent - when a large smoothing is used, the field is washed into the linear regime. The size of the CIC cells sets the smallest scale we may probe (i.e $0.125h^{-1}$Mpc), while in principle the largest scale is set simply by the simulation's box size.

{\bf Once the velocity field on a smoothed grid is obtained, the spatial derivatives are taken in Fourier space by means of an FFT. The symmetric (shear) and anti-symmetric (vorticity) components of the deformation tensor are then constructed according to eq.~\ref{eq:shear} and eq.~\ref{eq:vort}. The shear tensor is diagonalized and its eigenvalues and eigenvectors obtained.}

 \begin{figure}
 \includegraphics[width=20pc]{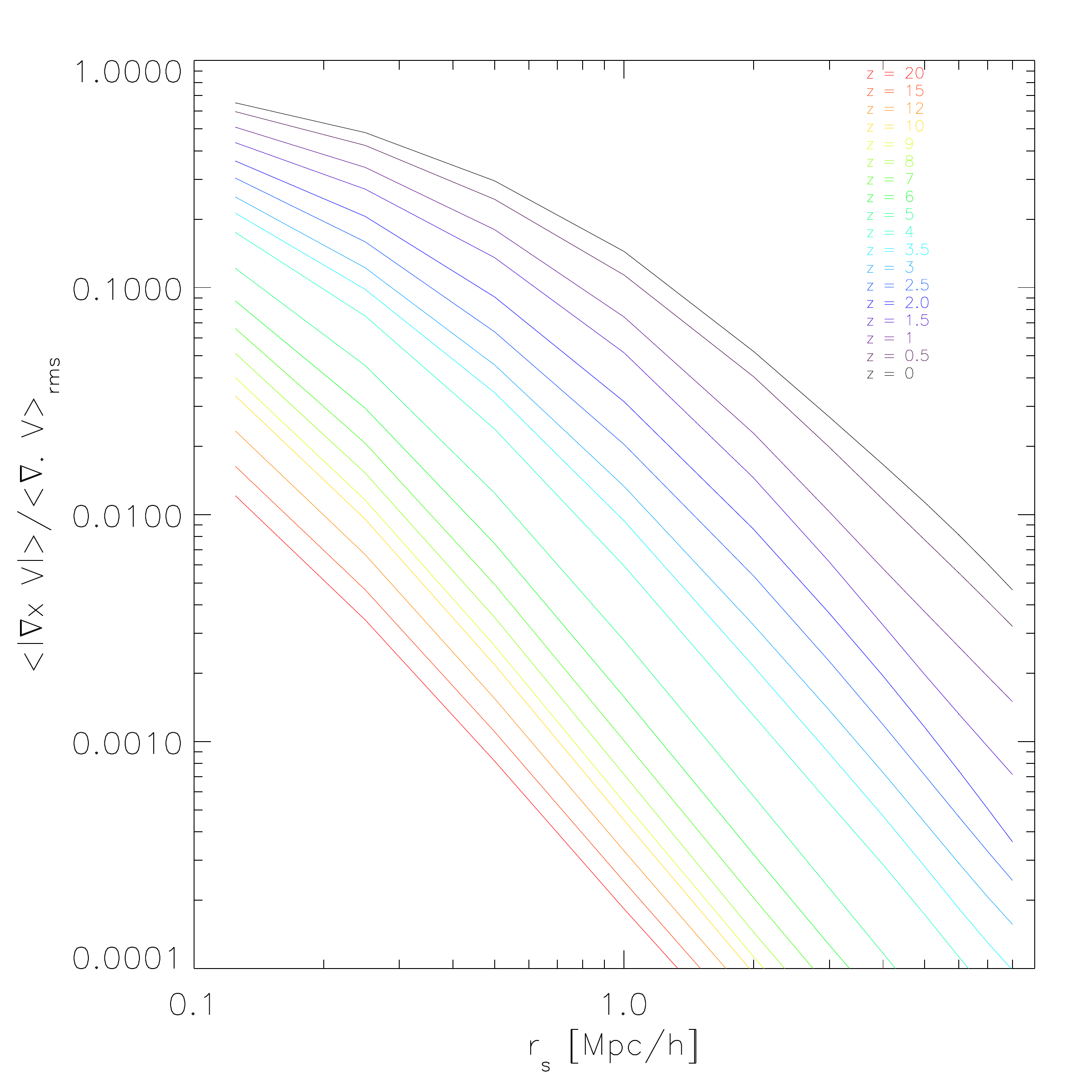}
\caption{The emergence of the vorticity as a function of smoothing scale and redshift. The magnitude of the vorticity is normalized to the magnitude of the divergence.}
\label{fig:emerge}
 \end{figure}

\section{Results}
 \label{section:results}

This section is devoted to the study of the evolution, in time and across spatial scales, of the symmetric and anti-symmetric components of the deformation tensor.  The different spatial scales are characterized by the application of a Gaussian kernel denoted by the smoothing scale, \rs. There is a symmetry when examining the universe across temporal and spatial scales: a journey across different smoothing scales is roughly equivalent to a journey through time. Changing the smoothing scale, from high to low \rs\ at a given redshift, corresponds to going from a less to a more dynamically evolved universe. It is thus similar to following the universe from high to low redshift at a fixed smoothing scale. One of the aims of the present study is to find an approximate scaling which uncovers the correspondence between spatial and temporal scales. The heart of the analysis is the study of the auto- and cross-alignment of the principal directions imposed by the deformation tensor, namely the three eigenvectors of the sheer tensor and the vorticity vector.

{\bf The vorticity is a strongly non-linear quantity and only emerges from non-negligible values for regions that are themselves non-linear. In Fig.~\ref{fig:magvortdens} we show the magnitude of the vorticity as a function of the local density, measured in units of the mean. In the quasi-linear regime the vorticity grows roughly proportional to the density ($\omega\propto\rho$).  Deep into the linear regime, at densities greater than roughly 10 times the mean, the vorticity roughly scales as $\rho^{0.3}$. It is important to note that for regions where $\rho \leq \rho_{\rm mean}$, the vorticity is small and essentially unimportant, dynamically speaking.}

In Fig.~\ref{figure:maps}, a thin slice of the $z=0$ snapshot is shown, smoothed with \rs$=0.125h^{-1}$ and 3.4$h^{-1}$Mpc on the left and right columns respectively. As mentioned above, $0.125h^{-1}$, is the highest resolution we may accurately probe, while $3.4h^{-1}$ is the scale of linearity, defined below (see \S~\ref{sec:rnl}).
The top row of Fig.~\ref{figure:maps} shows the density field with a blue contour separating regions that are above and below the mean density. In the bottom panel the importance of the vorticity is characterized by looking at its value relative to the magnitude of the divergence of the velocity, namely 
$|{\boldsymbol \omega}|/|\nabla \cdot {\bf v}|$.  Here the black contour denotes regions where this ratio is $ > 0.01$. At the highest smoothing most of the over-dense volume has a vorticity greater than 1\% of the divergence, highlighting the non-linear nature of vorticity generation. The greater smoothing of 3.4$h^{-1}$Mpc effectively erases both the high density peaks and the vorticity. Fig.~\ref{figure:maps} and Fig.~\ref{fig:magvortdens}
show that vorticity is a strong function of density and hence the scale on which it is measured. {\bf The following section examines the effect of the smoothing length on the density and velocity field. Since the scale of non-linearity increases with time we keep the ratio of the smoothing scale to 
the scale of non-linearity to be roughly constant (and equal to unity) in time, so as to be able to meaningfully compare the properties of the shear tensor at different time.}

\begin{figure*}
 \includegraphics[width=40pc]{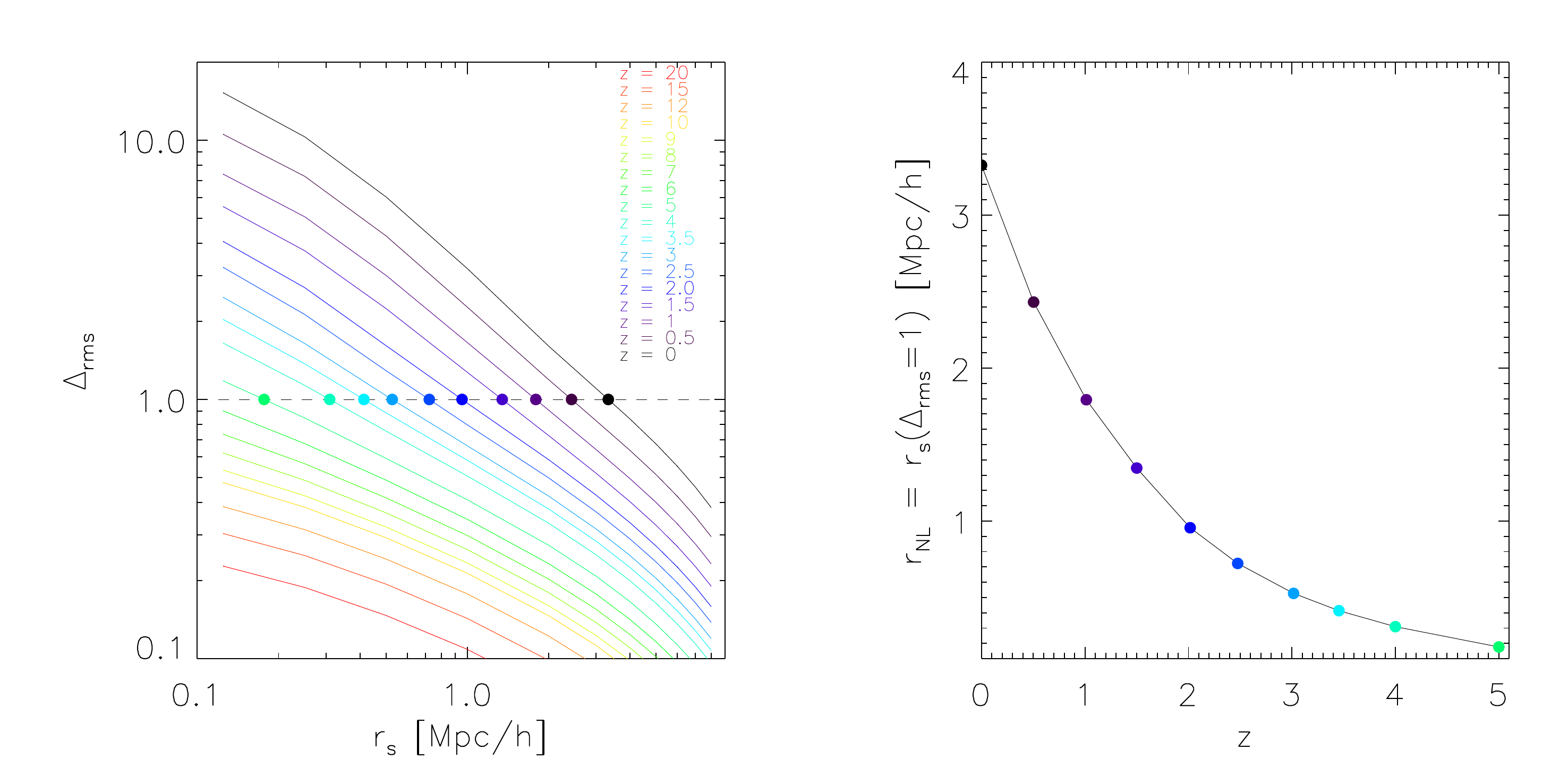}
 \caption{\textit{Left panel:} The {\it rms} of the density field (in units of the mean),  $\Delta_{\rm rms}$,   as function of smoothing scales for a variety of different redshifts. The smoothing scale for which $\Delta_{\rm rms}=1$ is taken as the ``scale of non-linearity'', \rnl. \textit{Right panel:} \rnl(z) is plotted as a function of redshift.}
\label{fig:smooths}
 \end{figure*}
\subsection{The effect of smoothing on the density and velocity field: scalar quantities}
\label{sec:effect}

Since the divergence of the velocity is, to some extent, a scalar measure of the symmetric part of the velocity deformation tensor, and because this changes more weakly with density than the anti-symmetric part, we expect the shear tensor to vary less with redshift and scale than the anti-symmetric part. The variability of the shear or the vorticity can be quantified either at a given redshift and as a function of scale or vice versa, namely at a given scale as a function of redshift. These two approaches are presented below.

We wish to quantify the subtle interplay between the symmetric and anti-symmetric tensors, as characterized by their relative magnitudes. The ratio 
$|{\boldsymbol  \omega}|/|\nabla \cdot {\bf v}|$ is shown as a function of smoothing for our fiducial redshifts in Fig.~\ref{fig:emerge}. At any given redshift the trend is uniform and expected: increasing the scale at which the deformation tensor is measured, decreases the importance of the vorticity. At $z=0$ the vorticity reaches a 1\% of the divergence on scales of $\sim5h^{-1}$Mpc. On Mpc scales the vorticity is significant, hinting that curl-free reconstructions of density and velocity fields on these scales (such as the Wiener filter algorithm of  \citealt{1995MNRAS.272..885F} and \citealt{1995ApJ...449..446Z}
and  the large body of work that came after these papers) might suffer from errors associated with assuming an irrotational flow.

\subsection{The scale of ``non-linearity''}
\label{sec:rnl}
 
In the previous section the importance of the vorticity with respect to the divergence was studied as a function of both redshift and scale (e.g. Fig.~\ref{fig:emerge}) . However, since it is asserted that vorticity is a non-linear effect, this scale must be defined as a function of redshift. In other words, \textit{what is the redshift evolution of the scale on which the density or velocity field is linear?} In order to address this issue we begin by trying to define what is nebulously meant by ``linear''.

One of the simplest statistical measures of the density field at each redshift is $\Delta_{\rm rms}$, the root mean square of the  normalized density field, $\Delta=\rho / \bar{\rho}$ where $\bar{\rho}$ is the mean density. One method to identify the scale of non-linearity, $r_{\rm NL}$, is to identify the smoothing scale which returns $\Delta_{\rm rms} =1$ at a given redshift.  
This procedure is shown in Fig.~\ref{fig:smooths}. The $\Delta_{\rm rms}$ calculated for a given redshift is plotted as a function of the smoothing scale. All redshift displays the same trend: the greater the smoothing, the lower $\Delta_{\rm rms}$, as expected. The scale on which these lines intersect unity (i.e. $\Delta_{\rm rms} =1$) is plotted as a function of redshift in the right panel of Fig.~\ref{fig:smooths}. The resolution of the simulation limits the redshift to $z<5$ on which the scale of $\Delta_{\rm rms}$ is found to be unity. That is, the smoothing scale for $z>5$ required to return $\Delta_{\rm rms}=1$ falls below the cells size of the CIC grid we have employed (which is determined by the simulations' resolution, see above).

Armed with a function $r_{\rm NL}(z)$ which returns the smoothing scale on which the $rms$ of the density field is unity, we may begin to probe how the anti-symmetric and symmetric part of the deformation tensor are aligned with each other.

 \begin{figure*}
 \begin{center}
 \includegraphics[width=42pc]{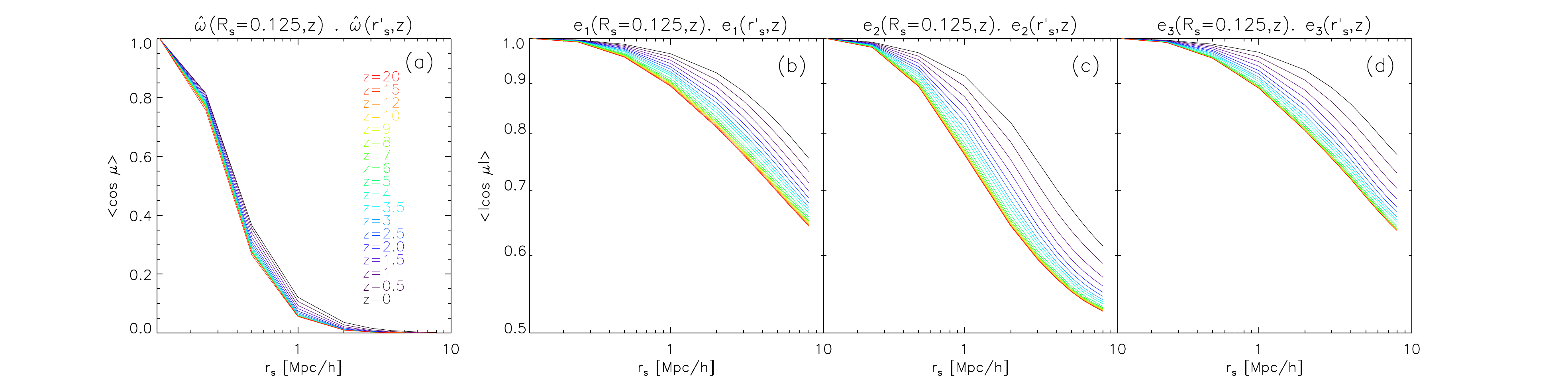}
 \caption{\textit{Left Panel:} (a) The median (cosine) of the angle, formed between the vorticity $\hat{\boldsymbol \omega}$ computed at $0.125h^{-1}$Mpc and as a function of smoothing scale (\rs) at a given redshift. A uniform distribution has a median $<\cos\mu>=0$. \textit{Right Panels:} the median (absolute value of the cosine) of the angle, formed between the principal axes of the shear (\eone, \etwo, and \ethree) computed at $0.125h^{-1}$Mpc and as a function of smoothing scale at a given redshift. Since the eigenvectors have no direction, a uniform distribution has a median $<|\cos\mu|>=0.5$. Note that the  vorticity is poorly aligned as a function of \rs, while the shear is more stable. Different redshifts are shown in different colors. }
\label{fig:shearcurlfr}
 \end{center}
 \end{figure*}

\subsection{Self alignment of the Shear and Vorticity tensor}

Much of this paper deals with the relative orientation of the various principal directions at varying smoothing scales and redshifts. These are described here in terms of the distribution of dot products between two vectors evaluated at a given redshift ($z$) and on a given scale (\rs), namely :
\begin{equation}
\label{ }
\cos\mu = {X}(r_{\rm s},z) \cdot {Y}(r{^{\prime}_{\rm s}},z^{\prime})
\end{equation}
Here $X, Y=$~\eone, \etwo, \ethree\  and ${\hat{\boldsymbol\omega}}$ are the (unit) vectors in question. Fixed smoothings and redshifts are denoted by unprimed letters (i.e. \rs\ and $z$); when the smoothing and/or the redshift is varied its denoted by a primed letter (i.e. $r{^{\prime}_{\rm s}},z^{\prime}$). For example, if the eigenvectors of the shear tensor (evaluated at a given redshift) are dotted with the vorticity (evaluated at the same redshift) at a fixed smoothing we write: ${\bf e}_{i}(r_{\rm s},z)\cdot\hat{\boldsymbol\omega}(r_{\rm s},z)$. If the eigenvectors of the shear tensor (evaluated at a given redshift and smoothing) are dotted with vorticity (evaluated at the same redshift but at a different smoothing) we write: ${\bf e}_{i}(r_{\rm s},z)\cdot\hat{\boldsymbol\omega}(r^{\prime}_{\rm s},z)$. We adopt the following convention: in the case that either $X$ or $Y$ are eigenvectors then it is the absolute value of the dot product that is taken (since eigenvectors are non-directional). The capital \Rs\ is reserved for the minimal smoothing length of $R_{\rm s}=0.125\hmpc$.

Section~\ref{sec:effect} emphasized that smoothing the velocity (and density) fields has a significant affect on the magnitude of quantities associated with the deformation tensor (namely $|\hat{\boldsymbol \omega}|$ and $\nabla\cdot {\bf v}$). Smoothing is also expected to have an affect on their principal directions, namely \eone, \etwo, \ethree, and 
${\hat{\boldsymbol\omega}}$.

{\bf In many of the figures that follow, median values for $\cos\mu$ computed from the full $512^{3}$ cells, are given. However, in cells which have very low densities or in regions which are still evolving linearly, the vorticity is a poorly defined quantity: its magnitude is tiny and its direction random. Furthermore, these cells can compose the majority of the simulation's volume at low $z$. In these cases, the unphysical (numerical) vorticity serves to weaken any alignment we find, which is driven by the most non-linear regions of the simulation. Therefore, most of the alignment plots shown in the following sections should be considered as lower limits to the alignment if it were to be measured only in regions where the vorticity is a physically well defined quantity.}

The effect of the smoothing scale on the orientation of the vorticity vector and the principal shear direction are shown in Fig.~\ref{fig:shearcurlfr}(a-d). 
The deformation tensor is calculated at smoothing lengths ranging from \Rs\ (i.e. $0.125\hmpc$) to $8.0\hmpc$, at redshifts ranging from 20 to the present epoch. The shear tensor eigenvectors and the vorticity are calculated at each cell in the full CIC grid.
Shown in Fig.~\ref{fig:shearcurlfr} is the median of the (cosine of this) angle at fixed $z$, namely ${\bf e}_{i}(R_{s},z)\cdot{\bf e}_{i}(r^{\prime}_{s},z)$ and ${\hat{\boldsymbol\omega}}(R_{s},z)\cdot{\hat{\boldsymbol\omega}}(r^{\prime}_{s},z)$. Note that for the vorticity, a median cosine of 0 is consistent with a random orientation (since $\hat{\boldsymbol\omega}$ is a vector and thus $\cos\mu$ ranges from [-1, 1] ), while for the shear a median cosine of 0.5 is consistent with a uniform random distribution (since shear eigenvectors are axes with no direction and thus $\cos\mu$ ranges from [0,1]).

Fig.~\ref{fig:shearcurlfr}(a), which shows the median of ${\hat{\boldsymbol\omega}}(R_{s},z)\cdot{\hat{\boldsymbol\omega}}(r^{\prime}_{s},z)$, indicates that as the smoothing scale 
increases, the orientation of the vorticity becomes random. Indeed this occurs relatively quickly: at a smoothing scale of just $r_{\rm s}\sim1~h^{-1}$Mpc, the vorticity for most redshifts has a median misalignment of $\approx 0.1$ (around 85$^{\circ}$) with respect to that calculated on a $0.125h^{-1}$Mpc scale. This is perhaps not surprising given that the vorticity on these scales is small ($\sim 10\%$ of the divergence at $z=0$, and $<1\%$ of the the divergence for z$>3.5$): the density field and vorticity magnitude have been smoothed out.

\begin{figure*}
\includegraphics[width=40pc]{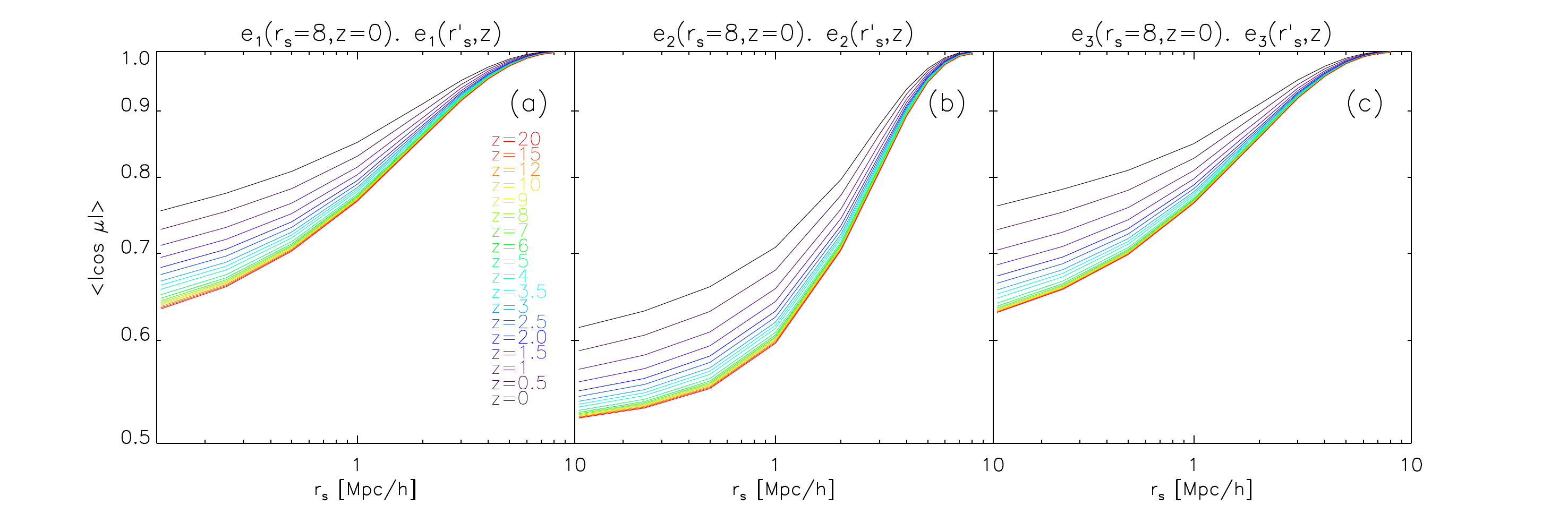}
\caption{
The median of  the cosine of the angle, formed between the shear eigenvectors  computed at a smoothing of  $r_s=8.0\hmpc$  and at $z=0$ and as a function of \rs and $z$.  A uniform distribution has a median $<|\cos\mu|>=0.5$. Different redshifts are shown in different colors. Left panel: median of ${e}_1(r^{\prime}_s, z)\cdot {e}_1(r_s=8.0\hmpc,z=0)$; Middle panel: same but for \etwo; Right panel: Same but for \ethree. 
}
\label{fig:shear-shear-cross}
 \end{figure*}

On the other hand, Figs.~\ref{fig:shearcurlfr}(b,c,d) show the median misalignment of the principal shear axes (\eone, \etwo, and \ethree) as a function of smoothing scale, at a given redshift, namely
${\bf e}_{i}(R_{s},z)\cdot{\bf e}_{i}(r^{\prime}_{s},z)$ (where $i=1,2,3$).
The variability here is weaker - large smoothing scales change the orientation of the shear eigenframe mildly. For example, at $z=20$ median misalignment of \eone\ or \ethree\ smoothed on 0.125 and on 1$h^{-1}$Mpc is just $\approx0.9$ (just 25$^{\circ}$). The median misalignment of \etwo\ on these two scales and at this redshift is larger, at around $\approx0.75$ ($\sim 40^{\circ}$)  These are our first results: although increasing the smoothing scale misaligns both the symmetric and anti-symmetric part of the deformation tensor, \textit{at a given redshift, the symmetric part is more robust to changes in scale while the anti-symmetric part, itself a strong function of density, is more fickle}.

Fig. \ref{fig:shear-shear-cross} presents the self-alignment of  the shear eigenvectors somewhat differently from Fig. \ref{fig:shearcurlfr}. It uses as the ``reference vector'' the shear eigenvectors smoothed on the $8.0h^{-1}$Mpc scale (instead of on \Rs) at $z=0$ and examines the angle between this and shear eignevectors calculated with {\it decreasing} smoothings at a fixed $z$, namely ${\bf e}_{i}(r_{\rm s}=8,z=0)\cdot{\bf e}_{i}(r^{\prime}_{s},z)$.  The rational behind the presentation of Fig. \ref{fig:shear-shear-cross} is that the shear tensor on the scale of $\approx 8h^{-1}$Mpc and at $z=0$ is accessible observationally and it can be estimated from full three dimensional  reconstruction of the velocity field \citep[e.g. the Wiener reconstruction of the velocity field from the Cosmicflows-1 data base,][]{2012ApJ...744...43C}. Also,  the gravitational tidal tensor can be extracted from galaxy redshift surveys and on the $8\hmpc$ scale it is closely aligned with the velocity shear tensor \citep[see, e.g.][]{LeeErdogdu2007}. Fig. \ref{fig:shear-shear-cross} shows the coherent orientation of the eigenvectors of the shear tensor from the observational accessible scales down to highly non-linear scales that are not currently feasible to observe. Note that just as seen previously, \eone\ and \ethree\ are well aligned across spatial scales  (Fig. \ref{fig:shear-shear-cross}a and Fig. \ref{fig:shear-shear-cross}c) while the alignment across scales for \etwo\ is weaker (Fig. \ref{fig:shear-shear-cross}b). Nevertheless none go to uniform and all eigenvectors of the shear tensor show coherence across Mpc scales.

\subsection{Cross alignment of the Shear and Vorticity tensor}
The "cross talk" between the shear and vorticity is examined here using three different approaches. 
In section \ref{sec:vortfix}, the vorticity is calculated on the smallest scales
and the angle between it and the shear is calculated as a function of scale and presented for many redshifts. In section~\ref{sec:nofix} both the shear and vorticity are calculated as a function of (the same) scale, and their alignment presented for each redshift. In section~\ref{sec:angrnl}, at each redshift, the shear and vorticity are calculated on scales determined by $r_{\rm NL}$ (see section~\ref{sec:rnl}).

\subsubsection{Cross alignment of vorticity and shear on the smallest scale}
\label{sec:vortfix}

\begin{figure*}
  \includegraphics[width=40pc]{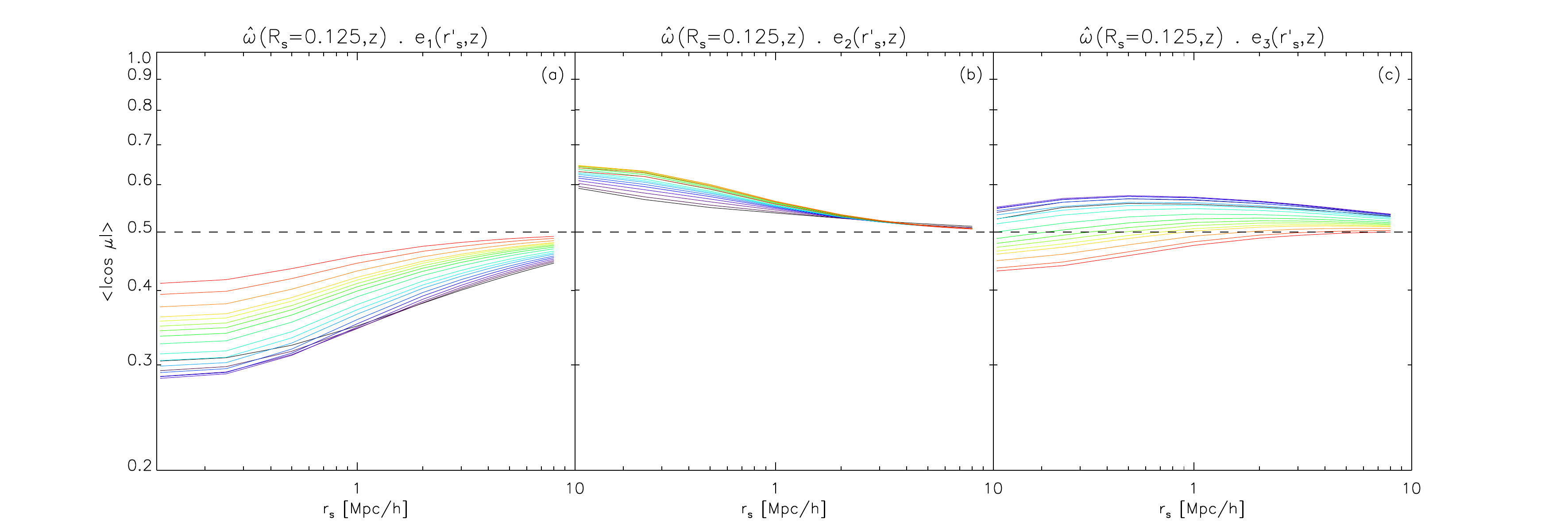}
\caption{The median of  the cosine of the angle formed between the vorticity computed at $0.125h^{-1}$Mpc and  \eone~(left), \etwo~(middle) and \ethree (right) as a function of smoothing scale (\rs) at a given $z$. A uniform distribution has a median $<|\cos\mu|>=0.5$ (dashed line). Different redshifts are shown in different colors. }
\label{fig:shearcurlcross}
 \end{figure*}

The shear tensor in general, and its principal directions in particular, are of primordial origin and are imprinted by the primordial perturbation field. Any primordial vorticity was completely damped out by the gravitational instability and existing present epoch vorticity has emerged out of the non-linear dynamics. This has prompted us to examine the  orientation of the vorticity relative to the eigenvectors of the shear tensor. This is done here by calculating the median of the alignment of smallest scale vorticity with the shear eigenvectors on variables scales. Fig.~\ref{fig:shearcurlcross} presents  the median of the $\hat{\boldsymbol\omega}(R_{\rm s},z)\cdot {\bf e}_{i}(r^{\prime}_{s},z)$, where  $R_{\rm s}=0.125\hmpc$ and  $i=1,2,3$.

A number of salient points are seen from these plots. Firstly, at high redshift and large smoothing scales ${\hat{\boldsymbol\omega}}$ approaches a uniform distribution (i.e. $|\cos\mu|\sim0.5$) with respect to all the shear eigenvectors. 
Fig.~\ref{fig:shearcurlcross}a shows that in the limit of the linear regime, i.e. large smoothing and early times, ${\hat{\boldsymbol\omega}}$ tends to be isotropically distributed relative to \eone. Moving away from the linear regime, i.e. small smoothing and late times,  ${\hat{\boldsymbol\omega}}$ tends to be more perpendicular to \eone. 
The distribution of ${\hat{\boldsymbol\omega}}$ with respect to \etwo\ follows a different trend (Fig.~\ref{fig:shearcurlcross}b). 
Irrespective of the redshift, the median of the distribution of  ${\hat{\boldsymbol\omega}}(R_{\rm s},z)\cdot{ \bf e}_2(r_{\rm s},z)$ is close to 0.5 for large smoothing scale. At small \rs\ the median is closer to 1, i.e. the vorticity tends to be aligned with \etwo, however as the universe evolves the vorticity vector moves away from \etwo\ and becomes less aligned with it. 
To the extent that our conjecture that the vorticity is associated with halo spin, then this is behavior is consistent with the argument made by  \cite{NavarroAbadiSteinmetz04} for the preferred alignment of ${\hat{\boldsymbol\omega}}$ with \etwo. The new result here is that this trend gets weaker, but does not disappear altogether, with time.
The cross alignment of ${\hat{\boldsymbol\omega}}$ and \ethree~displays even more subtle behavior (Fig.~\ref{fig:shearcurlcross}c). 
Considering \ethree, evaluated on the small scales, Fig.~\ref{fig:shearcurlcross}c shows that earlier times   ${\hat{\boldsymbol\omega}}$ shows a slight tendency to be perpendicular to it, and then with time it moves to be slightly aligned with it.

\subsubsection{Cross alignment of vorticity and shear at the same scale}
\label{sec:nofix}

\begin{figure*}
  \includegraphics[width=40pc]{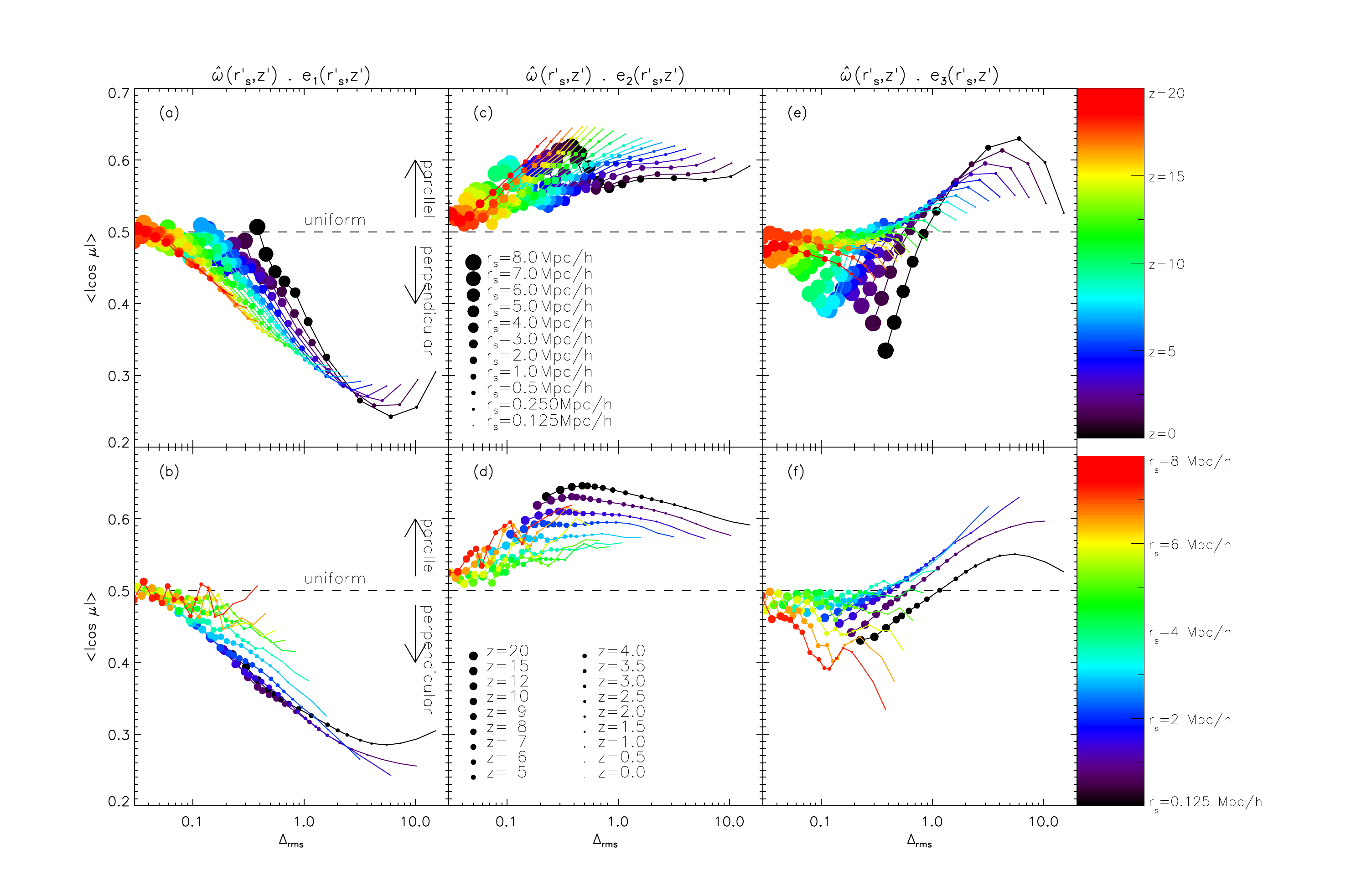}
 \caption{The median angle formed between $\hat{\boldsymbol\omega}$ and \eone~(left), \etwo~(middle), and \ethree~(right) as a function of $\Delta_{\rm rms}$. 
 Upper row: For each redshift (colored according to the color bar), the median for a given smoothing is indicated with point size proportional to the smoothing used. Smaller scales indicate a more perpendicular alignment. Bottom row: same as the top row but with symbols sized according to $z$ and connected by constant smoothing scales.}
 \label{fig:meddelta}
 \end{figure*}

Next, the alignment of the vorticity and the three shear eigenvector, evaluated at the same redshift and smoothing scale is studied here. Fig.~\ref{fig:meddelta} presents the median of the distribution of 
$ {\hat{\boldsymbol\omega}}(r_{\rm s},z)\cdot{\bf e}_{i}(r_{\rm s},z)$ ($i=1,2,3$). 
We argued before (\S \ref{sec:rnl}) that to a rough approximation the degeneracy between (\rs,\ z) pairs can be partially removed by 
considering the dependence of the alignment on $\Delta_{\rm rms}(r_{\rm s},z)$. This has driven us to plot the dependence of the alignment on $\Delta_{\rm rms}(r_{\rm s},z)$ in Fig.~\ref{fig:meddelta}.
The left, middle and right columns show the median of the cosine of the angle between the vorticity and \eone, \etwo, and \ethree\ respectively. The upper rows connect points at the same redshift, while the bottom rows connect points with the same smoothing length. As elsewhere in this work, the smoothing scale ranges between $0.125$ to $8.0\ \hmpc$. 

What emerges from Fig.~\ref{fig:meddelta} is that deep in the linear regime, smoothing scales and redshifts for which $\Delta_{\rm rms}  \lsim  0.1$, the orientation of the vorticity with respect to the three shear eigenvector is isotropic. This is manifested by the median of the (cosine) of the three angles being very close to 0.5. Moving away into the non-linear regime, namely towards higher values of $\Delta_{\rm rms}$, the vorticity moves away from \eone, and lies preferentially close to the (\etwo-\ethree) plane. The dynamics with that plane is more complex. The vorticity becomes initially more aligned with \etwo, but later on the distribution shifts to be more aligned with \ethree.   
Associating the vorticity with the spin of DM halos, then the late time distribution depicted here is consistent with Fig. 2 of \cite{2013ApJ...766L..15L}.

\subsubsection{Cross alignment of vorticity and shear on the scale of non-linearity}
\label{sec:angrnl}

\begin{figure*}
 \includegraphics[width=40pc]{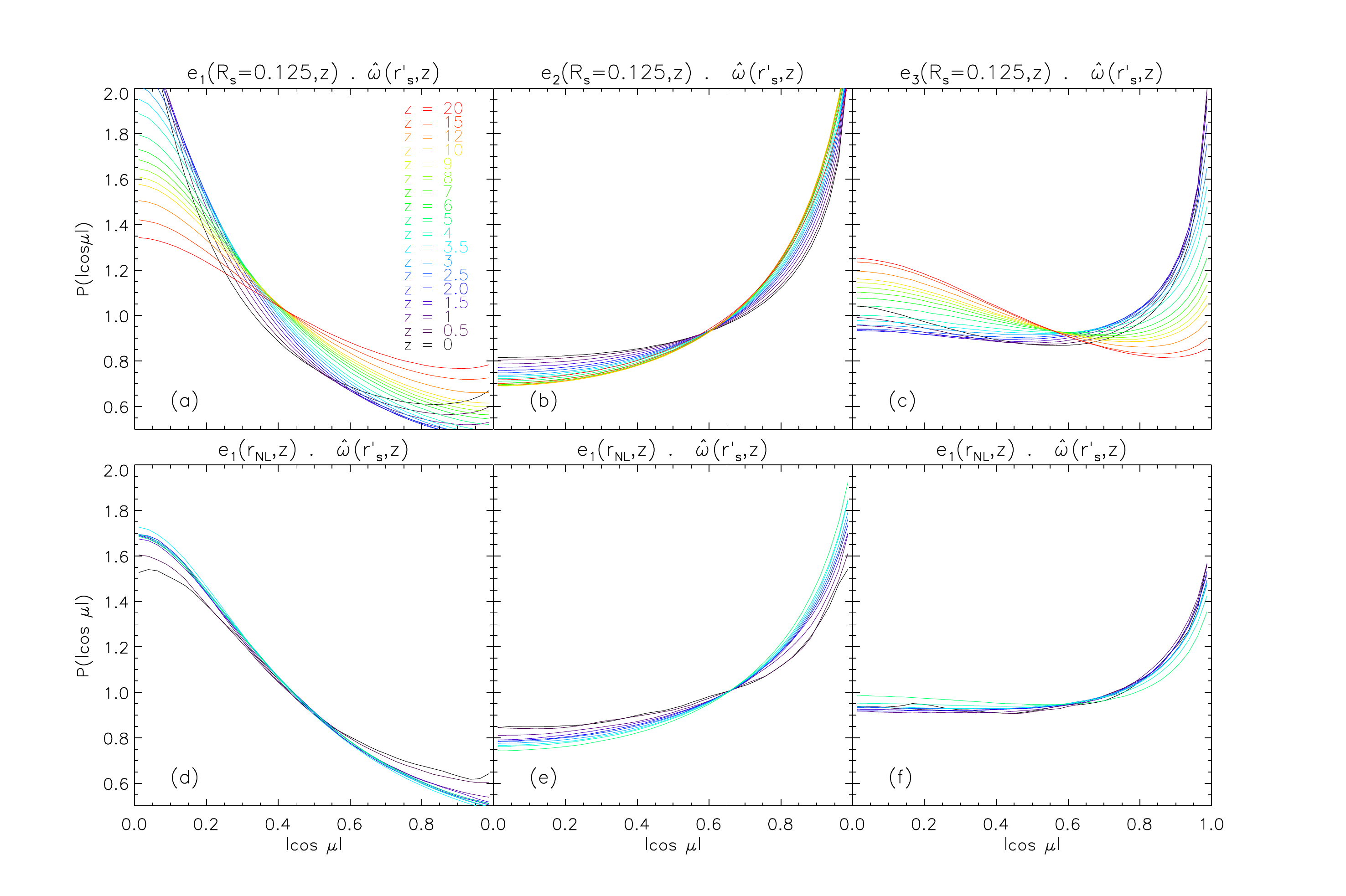}
 \caption{The probability distribution $P(|\cos\mu|)$, of the (cosine) of the angle formed between the vorticity $\hat{\boldsymbol\omega}$ and the principal axis of the shear (\eone\ -left, \etwo\ -middle, \ethree\ -right) for a variety of redshifts. A uniform distribution is a flat line equal to unity. \textit{Upper panels:} $P(| \cos\mu |)$ when smoothed on the highest  possible smoothing, $r_{\rm s}=0.125h^{-1}$Mpc. \textit{Bottom panels:} $P(| \cos\mu |)$) when smoothed, for each redshift with \rnl, defined in section~\ref{sec:rnl}. }
 \label{fig:vorte1align}
 \end{figure*}

The cross vorticity - shear alignment is calculated here on the scale of non-linearity, \rnl. Fig.~\ref{fig:vorte1align} shows the probability distribution of $\hat{\boldsymbol\omega}(r_{\rm NL},z)\cdot {\bf e}_{i}(r_{\rm NL},z)$ for a range of redshifts. 
The probability distribution of $\hat{\boldsymbol\omega}(r_{\rm NL},z)\cdot{\bf e}_{i}(R_{\rm s}, z)$ is shown for comparison (where $R_{\rm s}=0.125\hmpc$). Note that here the full distribution is presented, not only the median.

Although the limited dynamical range of the simulation limits the number of snapshots for which \rnl\ is resolved, Fig.~\ref{fig:vorte1align} shows that the distribution of the alignment with respect to the three eigenvectors is close to being redshift invariant. Namely, most of the dynamical evolution is absorbed by \rnl(z), which gives support to our choice of \rnl. 

The probability distribution of $\hat{\boldsymbol\omega}(r_{\rm NL},z)\cdot{\bf e}_{i}(R_{\rm s}, z)$, shown in  Fig.~\ref{fig:vorte1align}, supports the previous conclusions: non-linear evolution drives the vorticity to lie in the (\etwo-\ethree) plane, with a gradual change of alignment from \etwo\ to \ethree.

\section{Summary and Discussion}
\label{section:summary}

The large scale structure  (LSS) of the universe is often described in terms of the density field, namely by the distribution of mass in three dimensional configuration space. An alternative approach is to describe the LSS by means of the velocity field of the objects that constitute the density field. This 
approach consists of representing the particles' velocities by a continuous field evaluated on an arbitrary grid, much in the same way as the density field represents the mass distribution. 

Often the density field is used to characterize the small scale non-linear dynamics and evolution of the LSS. In this work, we do so by following the velocity deformation tensor which plays a major role in shaping the LSS of the universe, and in particular of the cosmic web. The deformation tensor also provides the principal directions along which halos and galaxies orient themselves. The deformation tensor can be decomposed into a symmetric and antisymmetric components, known as the velocity shear tensor and the vorticity. The symmetric part is imprinted in the initial conditions and grows throughout time in both the linear and non-linear regime. In the linear regime the velocity shear and the gravitational tidal tensors are, up to a normalization constant, identical. They start to depart in the non-linear regime. The curl of the velocity field, i.e. the vorticity, behaves very differently in the standard model of cosmology. Any primordial vorticity, if it had existed, would have been damped out in the linear phase of structure formation \citep[see][]{1980lssu.book.....P}. But shell crossing in the non-linear phase leads to the emergence of the vorticity \cite{1999A&A...343..663P}, which becomes closely related to the angular momentum of DM halos  \citep{2013ApJ...766L..15L}. This has motivated us to study the  time evolution  of the velocity deformation tensor and the auto- and cross-orientation of its symmetric and antisymmetric components across spatial and temporal scales. This is done by examining a high resolution DM-only cosmological simulations performed within the concordance cosmology, using the WMAP5 cosmological parameters.

One of the main results of this paper is the stability of the directions of the shear eigenvectors. Figure \ref{fig:shearcurlfr} shows the strong auto-alignment of \eone, the axis of fastest collapse, across different spatial scales. The plot shows the strong alignment between \eone, of the minimally smoothed velocity field, with that of the velocity field smoothed on scales extending up to $8.0 h^{-1}$Mpc. The strength of the alignment increases with time but it is already significant at $z=20$. This strong alignment of \eone\ across scales has an observational ramifications. It implies that from the observable shear tensor, on the $8h^{-1}$Mpc scale, one can predict the the orientation of \eone\ on sub-Megaparsec scales. Inferring the orientation of \eone\  on non-linear scales is important since the vorticity lies within the plane defined by \eone, namely the (\etwo-\ethree) plane. 

Arguably the  most significant result of this work is presented in Fig. \ref{fig:shearcurlcross}, which shows the relative orientation of the velocity curl, smoothed on the minimal scale probed in the paper, and the three velocity sheer eigenvectors smoothed on varying scale \rs. These orientations are evaluated all the way back to $z=20$ and out to a smoothing scale \rs$=8.0 \hmpc$. The orientation of the vorticity with \eone\ and \etwo\ evolves as expected. Already at the onset of the non-linear regime, at large smoothing and/or large redshift, the vorticity tends to be orthogonal to \eone. Within the (\etwo-\ethree) plane, the vorticity starts aligned with \etwo\ and then moves towards \ethree.  To the extent that the vorticity is associated with the angular momentum of DM halos, then the orientation of the ${\hat{\boldsymbol\omega}}$ with respect to \eone\ and \etwo\ remains consistent with the predictions of Tidal Torque Theory (TTT), with the non-linear dynamics just strengthening the linear trends. The evolution with respect to \ethree\ starts as predicted by TTT, but then the dynamical evolution drives the two vectors towards being even more parallel. The distribution of the vorticity with respect to the three shear eigenvector  is consistent with the finding of \cite{2013ApJ...766L..15L} concerning the redshift zero orientation of the halos' spin with respect to the eigenvector. The spins and the vorticity are distributed alike. 

The \etwo-\ethree\ flipping is intriguing and likely reflective of the angular momentum - \ethree\ flip seen for high mass haloes in \cite{2013MNRAS.428.2489L} and \cite{2012MNRAS.427.3320C}: at $z\lsim2$ the vorticity tends towards being parallel to \ethree~for a given physical scale of \rs$\approx0.5h^{-1}$Mpc. Below or above that scale the alignment randomizes. This is remarkably similar to the scale that corresponds to the transition mass found by \cite{2013MNRAS.428.2489L} and \cite{2012MNRAS.427.3320C}.

The vorticity vector is aligned with itself only on sub-Mpc scales where it is well defined. Namely, the median of  ${\hat{\boldsymbol\omega}}(R_{\rm s},z)\cdot\hat{\boldsymbol\omega}(r_{\rm s},z)$ is larger than $0.5$ for all redshift concerned here only for $r_{\rm s} \lsim 0.4$. This is very close to spin-spin alignment correlation length of DM halos 
\citep[][and references therein]{2012arXiv1201.6108T}. This close agreement between the alignment of the vorticity field with itself and of the spin of DM halos provides further support for our conjecture between the close correspondence of the curl of the velocity field and the spin of DM halos, and by association also the spin of galaxies.

Much of the interest in studying the cosmic web is driven by the fact that it dictates preferred directions along which DM halos, and thereby galaxies, orient themselves. However, progress is hindered here by the fact that both halos and the cosmic web are not clearly and unequivocally defined objects. The multitude of definitions are a testimony of the ambiguity inherent in the construction of these objects. \cite{2013MNRAS.428.2489L}  and this work suggest a new framework, in which the cosmic web is represented by the eigenvectors of the shear tensor and the spin of DM halos by the vorticity. 
This has motivated us to focus here on the differential properties of the velocity field, in particular on the symmetric and antisymmetric components of the deformation tensor. 
The extension of the present work to DM halos and the cosmic web,  across different spatial and temporal scales will be presented in a forthcoming paper.

While writing this paper two related articles appeared on the arXiv: \cite{2013arXiv1309.5305W} and \cite{2013arXiv1310.3801L}. \cite{2013arXiv1309.5305W} have posted a preprint which presents a study of the evolution of the cosmological velocity from potential to rotational flow. The analytical approach of Wang et al complements the numerical analysis of \cite{2013MNRAS.428.2489L} and the present paper. \cite{2013arXiv1310.3801L} also examined the results of  \cite{2013MNRAS.428.2489L} in great detail scrutinizing how the halo spin-vorticity alignment is assembled. Together these papers suggest a new framework for the investigation of the mergence of structure out of an expanding Friedmann universe via gravitational instability.

\section*{Acknowledgments}
NIL is supported by the Deutsche Forschungs Gemeinschaft, YH has been partially supported by the Israel Science Foundation (1013/12). SG and YH have been partially supported by the Deutsche Forschungsgemeinschaft under the grant $\rm{GO}563/21-1$. The simulations have been performed at the Leibniz Rechenzentrum (LRZ) in Munich.

\bibliography{./ref}
\end{document}